\documentclass[twocolumn]{webofc}

\usepackage[varg]{txfonts}   
\newcommand{\be}{\begin{equation}}
\newcommand{\ee}{\end{equation}}
\newcommand{\bea}{\begin{eqnarray}}
\newcommand{\eea}{\end{eqnarray}}
\begin{document}
\title{Nuclear response at zero and finite temperature}
%
%

\author{\firstname{Elena} \lastname{Litvinova}\inst{1,2}\fnsep\thanks{\email{elena.litvinova@wmich.edu}} \and
        \firstname{Peter} \lastname{Schuck}\inst{3,4}\fnsep\thanks{\email{schuck@ipno.in2p3.fr}} \and
        \firstname{Herlik} \lastname{Wibowo}\inst{1}\fnsep\thanks{\email{herlik.wibowo@wmich.edu}}
}

\institute{Department of Physics, Western Michigan University, Kalamazoo, MI 49008, USA
\and
           National Superconducting Cyclotron Laboratory, Michigan State University, East Lansing, MI 48824, USA 
\and
           Institut de Physique Nucl\'eaire, IN2P3-CNRS, Universit\'e Paris-Sud, F-91406 Orsay Cedex, France
\and
           Universit\'e Grenoble Alpes, CNRS, LPMMC, 38000 Grenoble, France
          }
\abstract{%
 We present some recent developments on the nuclear many-body problem, such as the treatment of high-order correlations and finite temperature in the description of in-medium two-nucleon propagators.  In this work we discuss two-time propagators of the particle-hole type, which describe the response of finite nuclei to external probes without nucleon transfer. The general theory is formulated in terms of the equation of motion method for these propagators with the only input from the bare nucleon-nucleon interaction. 
The numerical implementation was performed on the basis of the effective mason-nucleon Lagrangian in order to study the  energy-dependent kernels of different complexity. The finite-temperature extension of the theory with $ph\otimes phonon$ configurations is applied to a study of the multipole response of medium-mass nuclei. }
\maketitle
%

{\it Introduction \textemdash} Response of many-body quantum systems to external perturbations comprises a large class of problems studied in many subfields of quantum physics. The observed spectral distributions are uniquely associated with quantum  
correlation functions, which encrypt the entire information about strongly-correlated complex media. 
Response of atomic nuclei to various experimental probes represents a very rich playground to study correlation functions, which was further extended with the advent of rare beam facilities \cite{Tanihata1998,Glasmacher2017}.  Nuclear correlation functions have a long history of theoretical studies based on the quantum field theory (QFT) techniques.  For instance, the idea of coupling between single-particle and emergent collective degrees of freedom in nuclei \cite{BohrMottelson1969,BertschBortignonBroglia1983,Soloviev1992}, which explained successfully many of the observed phenomena, can be related to the non-perturbative versions of QFT-based and, in principle, exact, equations of motion (EOM) for correlation functions in nuclear medium \cite{Rowe1968,Schuck1976,AdachiSchuck1989,DukelskyRoepkeSchuck1998,Storozhenko2003,SchuckTohyama2016}. In this work we discuss how a hierarchy of non-perturbative approximations to the dynamical kernels of the EOM's for one-fermion and two-time two-fermion propagators
can be mapped to the kernels of the phenomenological nuclear field theories (NFT), where such kernels are commonly referred to as particle-vibration coupling (PVC) or quasiparticle-phonon (QPM) models. This mapping
provides an understanding of the emergent collective phenomena, explaining the mechanism of their formation from the underlying strongly-interacting degrees of freedom.  Last decade, the latter type of approaches was linked to the contemporary density functional theories \cite{LitvinovaRing2006,LitvinovaRingTselyaev2007,LitvinovaRingTselyaev2008,LitvinovaRingTselyaev2010,LitvinovaRingTselyaev2013,AfanasjevLitvinova2015,Litvinova2016,RobinLitvinova2016,RobinLitvinova2018,LitvinovaRingTselyaev2010,Tselyaev2013,Tselyaev2016,NiuNiuColoEtAl2015,Niu2018}, advancing the PVC models to self-consistent frameworks and providing a better understanding of the low-energy nuclear structure \cite{EndresLitvinovaSavranEtAl2010,MassarczykSchwengnerDoenauEtAl2012,PoltoratskaFearickKrumbholzEtAl2014,LanzaVitturiLitvinovaEtAl2014,Oezel-TashenovEndersLenskeEtAl2014}. However, very little progress has been made on the conceptual advancements of the many-body aspects of the non-perturbative NFT's, except for some specific rare topics, such as PVC with charge-exchange phonons \cite{Litvinova2016,RobinLitvinova2018} or PVC-induced ground state correlations \cite{Kamerdzhiev1991,Robin2019}. More specifically, a relatively little effort was made on developments and numerical implementations of NFT's beyond the two-particle-two-hole $2p-2h$ level \cite{Tselyaev2018,Litvinova2015}, although the phenomenological multiphonon approach \cite{Soloviev1992,Ponomarev2001,Andreozzi2008}  indicates the possibility to meet the shell-model standards in large model spaces. Also, the problem of consistent linking those approaches with the underlying bare interactions remains unsolved. 
In the present work we revise the past and recent developments of the EOM method and  compare them to the PVC approach in the time blocking approximation (TBA)\cite{Tselyaev1989,LitvinovaRingTselyaev2007}. As an outcome, we formulate an extended strongly-coupled theory, which includes higher-order configurations with emergent long-range correlations corresponding to collective degrees of freedom. Moreover, we set a prospect for a consistent and systematically improvable approach to fermionic correlation functions, which is directly linked to the bare two-fermion interaction. The presented numerical implementations are, however, still based on an effective interaction, which allows us to test the long-range correlation sector of the theory at zero and finite temperatures. 
%

{\it Formalism and calculations \textemdash}
We consider a system of fermions with the Hamiltonian\footnote{We elaborate here on the detailed equation of motion based on this general Hamiltonian, which is most commonly used in nuclear physics. The generalizations to multinucleon Hamiltonians, relativity and adding explicit bosonic degrees of freedom are straightforward.}:
\vspace{5mm}
\be
H = \sum_{12}{t}_{12}{\psi^{\dagger}}_1\psi_2 + \frac{1}{4}\sum\limits_{1234}{\bar v}_{1234}{\psi^{\dagger}}_1{\psi^{\dagger}}_2\psi_4\psi_3 = T + U,
\ee
where the matrix elements of kinetic energy $t_{12} =  \delta_{12}\varepsilon_1$ and antisymmetrized interaction ${\bar v}_{1234} = v_{1234} - v_{1243}$ are given in the basis which diagonalizes $t_{12}$.
The number subscripts  '1' represent the full set of the one-fermion quantum numbers in this basis. 
The operators $\psi(1),{\psi^{\dagger}}(1)$ are fermionic (for instance, nucleonic) fields in the Heisenberg picture:
\be
\psi(1) = e^{iHt_1}\psi_1e^{-iHt_1}, \ \ \ \ \ \ {\psi^{\dagger}}(1) = e^{iHt_1}{\psi^{\dagger}}_1e^{-iHt_1},
\ee
The excitation spectrum of such a system is characterized by the two-time particle-hole propagator (response function): 
\be
R(12,1'2') \equiv R_{12,1'2'}(t-t')  = -i\langle T(\psi^{\dagger}_1\psi_2)(t)(\psi^{\dagger}_{2'}\psi_{1'})(t')\rangle = \nonumber 
\ee
\be
= -i\langle T\psi^{\dagger}(1)\psi(2)\psi^{\dagger}(2')\psi(1')\rangle,
\label{phgf}
\ee
assuming that $t_1 = t_2 = t, t_{1'} = t_{2'} = t'$ and $T$ is the chronological ordering operator,
while the averaging in Eq. (\ref{phgf}) $\langle ... \rangle$ is performed over the ground state in the case of $T=0$ and over the grand canonical ensemble at $T>0$.

\begin{figure}[h]
\centering
\sidecaption
\includegraphics[width=6cm,clip]{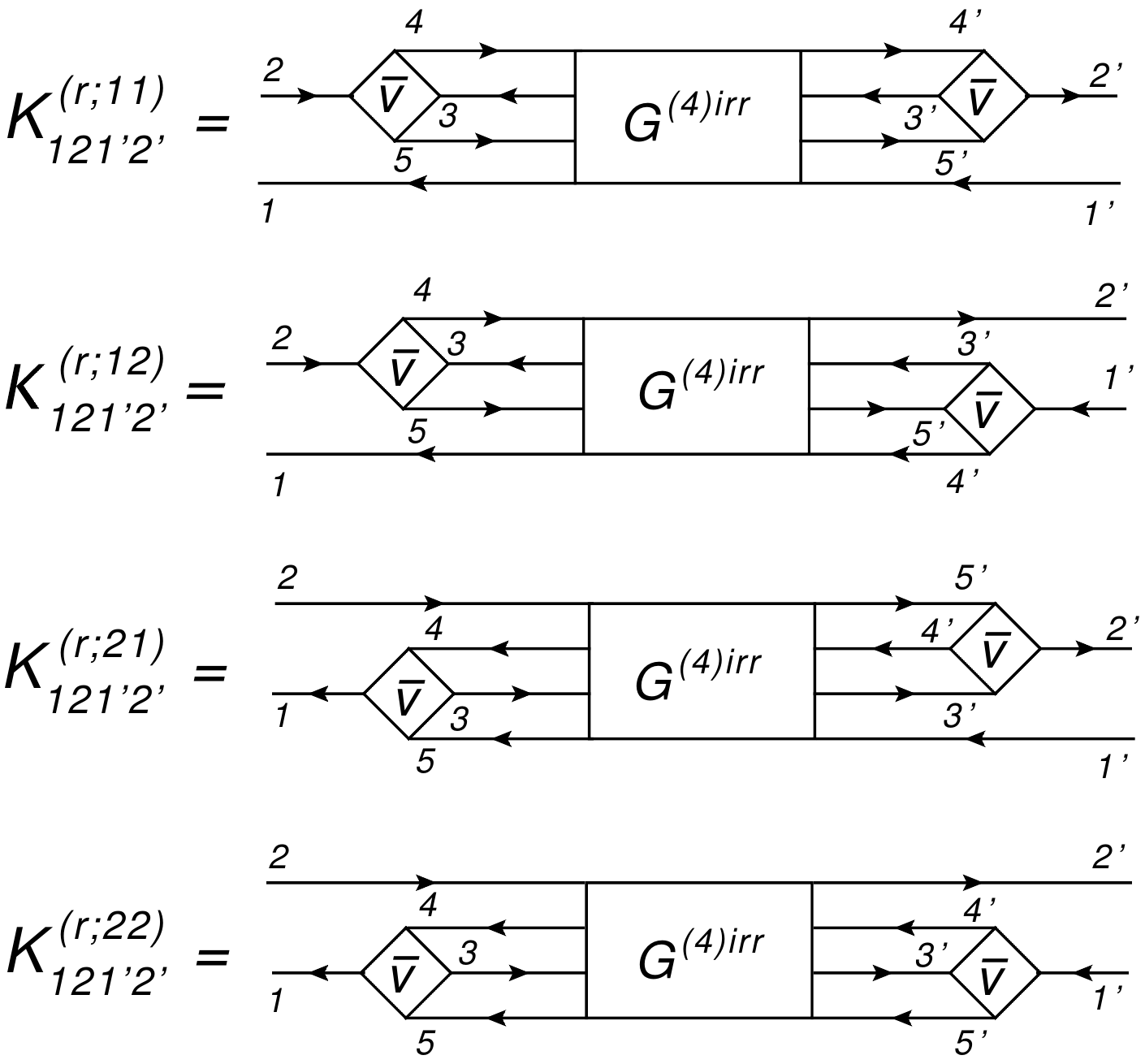}
\caption{The components of the dynamical kernel $K^{(r)}_{12,1'2'}(t-t')$. Straight solid lines stand for fermionic propagators, the square blocks denote the antisymmetrized nucleon-nucleon interaction $\bar v$, and the rectangular blocks $G^{(4)}$ correspond to the two-particle-two-hole Green function (\ref{G4}).}
\label{fig-1}       
\end{figure}

First we discuss the case of zero temperature. Differentiation of Eq. (\ref{phgf}) with respect to the time arguments generates the equation of motion for $R_{12,1'2'}(t-t')$, which takes the form of the Dyson equation 
after the Fourier transformation to the energy domain:
\be
R(\omega) = R^{(0)}(\omega) + R^{(0)}(\omega)K(\omega)R(\omega).
\label {Dyson2}
\ee
The $R^{(0)}(\omega)$ denotes the amplitude of particle-hole propagation without interaction (uncorrelated propagator):
\be
R^{(0)}_{12,1'2'}(\omega) = \frac{{\cal N}_{121'2'}}{\omega - \varepsilon_{2} + \varepsilon_{1}}, \ \ \ \ \ \ \ \ {\cal N}_{121'2'}=\delta_{11'}\delta_{22'}{\cal N}_{12},
\ee
where ${\cal N}_{12} = n_1 - n_2$ with $n_1 = \langle{\psi^{\dagger}}_1\psi_1\rangle$ being the occupation number of the single-particle state $1$.
The interaction kernel $K(\omega)$ of Eq. (\ref{Dyson2}) is  the amplitude
\be
K_{12,1'2'}(t-t') = \Bigl[-\delta(t-t')
\langle [[V,{\psi^{\dagger}}_1\psi_2],{\psi^{\dagger}}_{2'}\psi_{1'}]\rangle + \ \ \ \ \ \ \nonumber 
\ee
\be
+ i\langle T[V,{\psi^{\dagger}}_1\psi_2](t)[V,{\psi^{\dagger}}_{2'}\psi_{1'}](t')\rangle^{irr}\Bigr]{\cal N}_{12}^{-1}{\cal N}_{1'2'}^{-1}, 
\label{F1}
\ee
irreducible in the particle-hole channel and
Fourier-transformed to the energy domain.
%
It is clearly split into the instantaneous $K^{(0)}$ and the time-dependent $K^{(r)}$ parts:
$K(t-t') = K^{(0)}\delta(t-t') + K^{(r)}(t-t')$.
The instantaneous term represents the self-consistent phonon-mean field which is, thereby, derived from the bare interaction and contains two-body fermionic density being the equal-times limit of the particle-hole response function (\ref{phgf}) \cite{Olevano2018,SchuckTohyama2016a}.
The time-dependent part $K^{(r)}$ of the kernel $K$ is given in the diagrammatic form in Fig. \ref{fig-1}, 
where $K^{(r)} = \sum_{ij}K^{(r;ij)}$, the subscripts $i,j$ take the values of 1 and 2,  and we omitted constant factors  $\pm i/4$ in front of each term \cite{LitvinovaSchuck2019}. All of them include the two-time two-particle-two-hole (4-fermion) Green function $G^{(4)}$ defined as:
\bea
G^{(4)}(543'1',5'4'31) = \nonumber \\ = \langle T({\psi^{\dagger}}_1{\psi^{\dagger}}_3\psi_5\psi_4)(t)
({\psi^{\dagger}}_{4'}{\psi^{\dagger}}_{5'}\psi_{3'}\psi_{1'})(t')\rangle. 
\label{G4}
\eea
%
Evaluating this propagator is the central problem for the description of the dynamical kernel beyond the lowest-order with respect to the interaction ${\bar v}$, that is essential for strongy-coupled systems.  The direct EOM for this propagator generates higher-rank propagators, resulting in an infinite sequence of EOM's. Alternatively, 
efficient truncation schemes have been suggested, for instance,  in Refs. \cite{SchuckTohyama2016,SchuckTohyama2016a,Olevano2018}, where $G^{(4)}$ is treated as a superposition of products of two-fermion particle-hole $(ph)$ correlation functions (\ref{phgf}) and particle-particle $(pp)$ ones. As discussed in detail in Ref. \cite{LitvinovaSchuck2019}, in this non-perturbative approximation the problem reduces to the closed system of equations:
\be
{\hat R}(\omega) = {\hat R}^{(0)}(\omega) + {\hat R}^{(0)}(\omega)K[{\hat R}(\omega)]{\hat R}(\omega),
\label{Dyson3}
\ee
where 
${\hat R} = \Bigl\{ R^{(ph)}, R^{(hp)}, R^{(pp)}, R^{(hh)} \Bigr\}$ and the same structure is implied for the uncorrelated propagators ${\hat R}^{(0)}$.  
%

\begin{figure}
\centering
\includegraphics[width=6cm,clip]{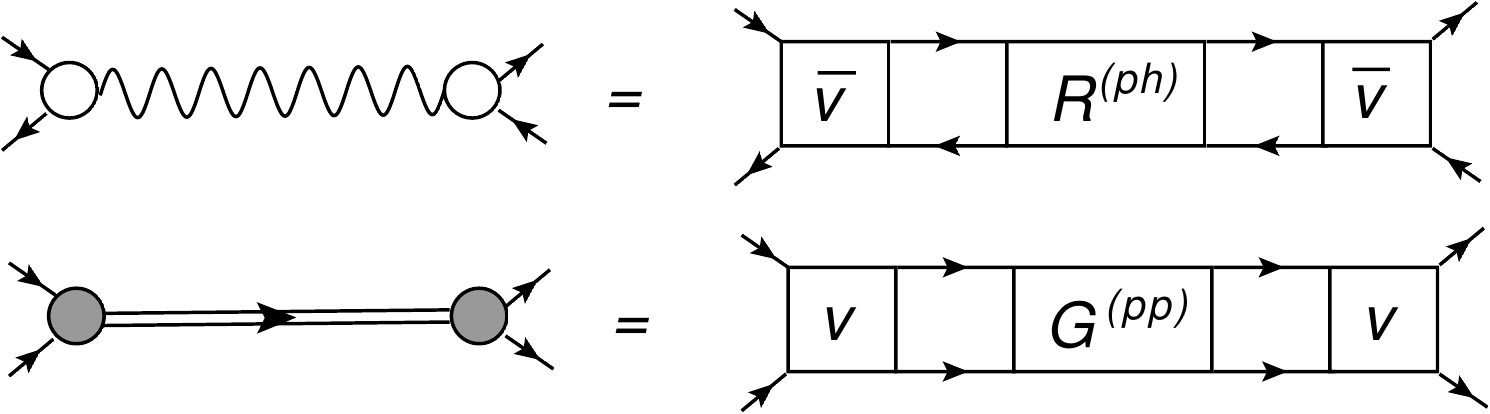}
\caption{Diagrammatic mapping of the EOM  to the PVC-TBA. Empty and filled circles denote the coupling vertices of the normal and pairing phonons, respectively, and the wiggly and double lines their propagators. $R^{(ph)}$ and $G^{(pp)}$ are the particle-hole response function and the particle-particle Green function, respectively. }
\label{fig-2}       
\end{figure}

\begin{figure}
\vspace{-2mm}
\centering
\includegraphics[width=8.2cm,clip]{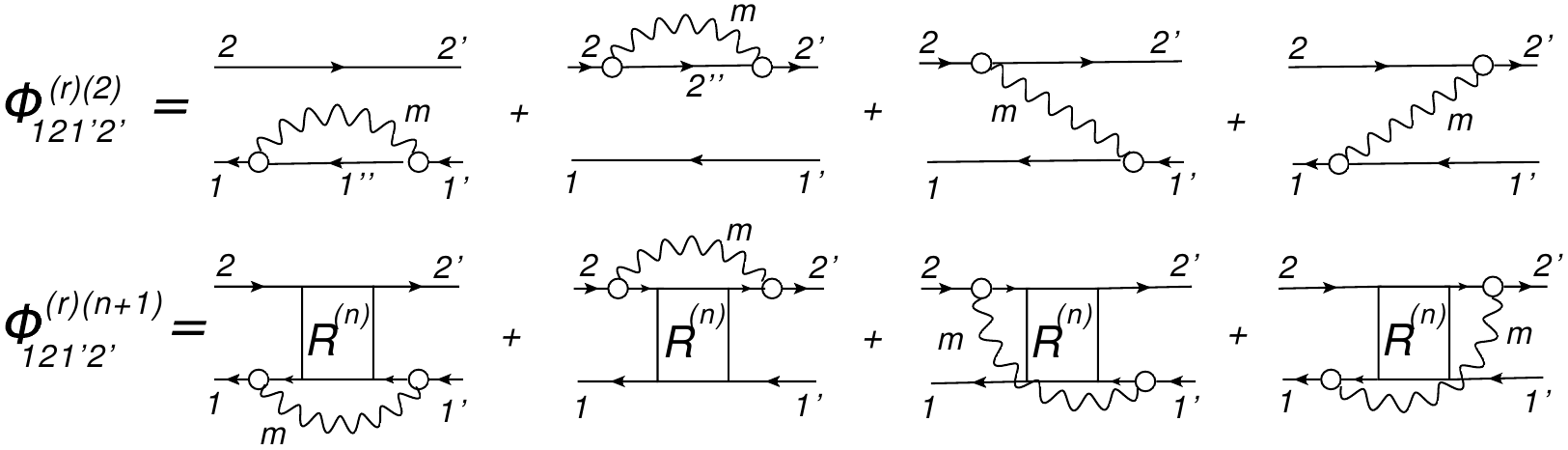}
\caption{Dynamical kernel of the conventional PVC-TBA (top) and generalized  EOM-TBA (bottom). }
\vspace{-5mm}
\label{fig-3}       
\end{figure}

To solve Eq. (\ref{Dyson3}) in practice, one needs a reasonable initial approximation for ${\hat R}$ to generate a convergent iterative procedure. %
This turns out to be possible in approaches based on the nuclear energy density functionals, which are adjusted to approximate the static part $K^{(0)}$ of the kernel $K$ by the effective interaction. In this case the initial ${\hat R}$ can be computed with only the static part of the integral kernel of Eq. (\ref{F1}), i.e. in the random phase approximation (RPA). Thus, reasonably good results were obtained in the relativistic quasiparticle time blocking approximation (RQTBA), where this static part is described by the exchange of effective mesons whose coupling constants and masses are adjusted to the nuclear ground state properties on the Hartree level \cite{VretenarAfanasjevLalazissisEtAl2005}. The dynamical part of the kernel was modeled by the PVC in the lowest order with respect to the phonon coupling vertex, following the non-relativistic (quasiparticle) time blocking approximation ((Q)TBA) \cite{Tselyaev1989,LitvinovaTselyaev2007,Tselyaev2007}. As we show in Ref. \cite{LitvinovaSchuck2019}, the (RQ)TBA kernel is topologically equivalent to the part of the EOM kernel with a single two-fermion correlation function while another pair of fermions remains uncorrelated. This equivalence was found by performing the mapping illustrated diagrammatically in Fig. \ref{fig-2} and the resulting (RQ)TBA kernel is displayed in the upper part of Fig. \ref{fig-3}.
The topological equivalence means that in the kernel of the (RQ)TBA approach based on the effective interaction this interaction replaces the bare interaction of the EOM while the one-fermion propagators become the mean-field propagators. Notice that the use of the effective interaction provides a good convergence of the iterations - in practice, one iteration is usually sufficient.
%
%

\begin{figure}
\centering
\includegraphics[width=7cm,clip]{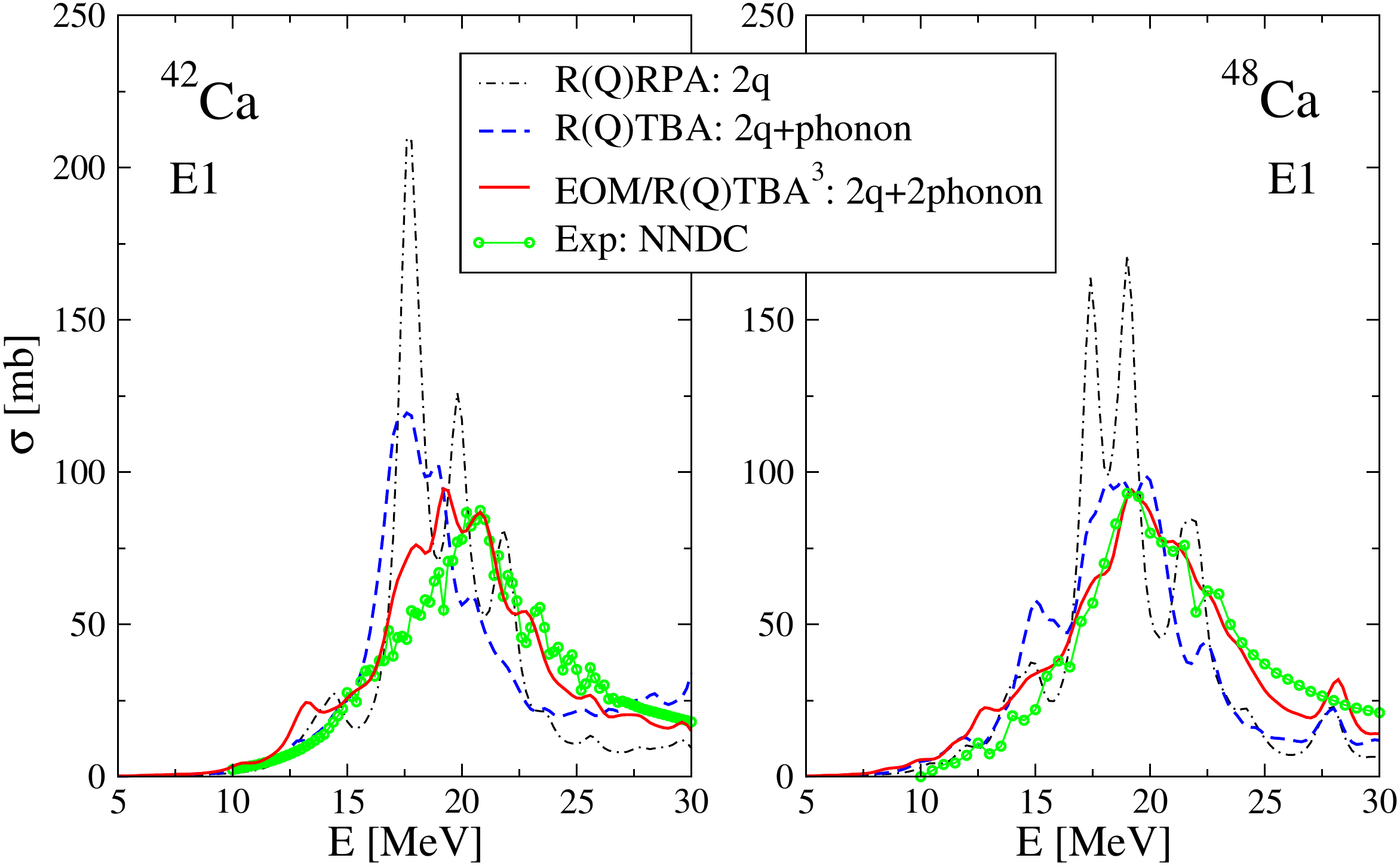}
\caption{Theoretical dipole photoabsorption cross sections in $^{42,48}$Ca compared to the evaluated experimental data of Refs. \cite{erokhova2003,nndc}.}
\label{fig-4}       
\end{figure}

In this work we follow the EOM method and combine it with the relativistic nuclear field theory (RNFT). The corresponding dynamical kernel of the new approach named EOM/RQTBA$^3$ is given in the lower part of Fig. \ref{fig-3}, where the index $"(n)"$ denotes the iteration order. The numerical implementation has been performed for $n=2$ based on the effective NL3 meson-nucleon interaction \cite{Lalazissis1997a} of quantum hadrodynamics (QHD), i.e. the static part of the interaction kernel is still approximated by the effective meson-exchange interaction. The second correlation function introduced into the dynamical kernel leads to an additional fragmentation of the strength distributions. The results of calculations for the giant dipole resonance (GDR) in $^{42,48}$Ca are displayed in Fig. \ref{fig-4} in comparison to data. These cross sections were studied  in Ref. \cite{EgorovaLitvinova2016} for the role of the $2q\otimes phonon$ (two quasiparticles coupled to a phonon) configurations in the formation of the spreading width, which was found quite significant as compared to relativistic quasiparticle random phase approximation (RQRPA) with only $2q$ (two-quasiparticle) configurations. However, the total width and the high-energy strength obtained in RQTBA are still insufficient to reproduce the data. Now we can see that EOM/RQTBA$^3$ with more complex $2q\otimes 2phonon$ configurations shows the potential of resolving those problems by inducing a stronger fragmentation of the GDR and its additional spreading toward both higher and lower energies. 
\begin{figure}
\centering
\includegraphics[width=7cm,clip]{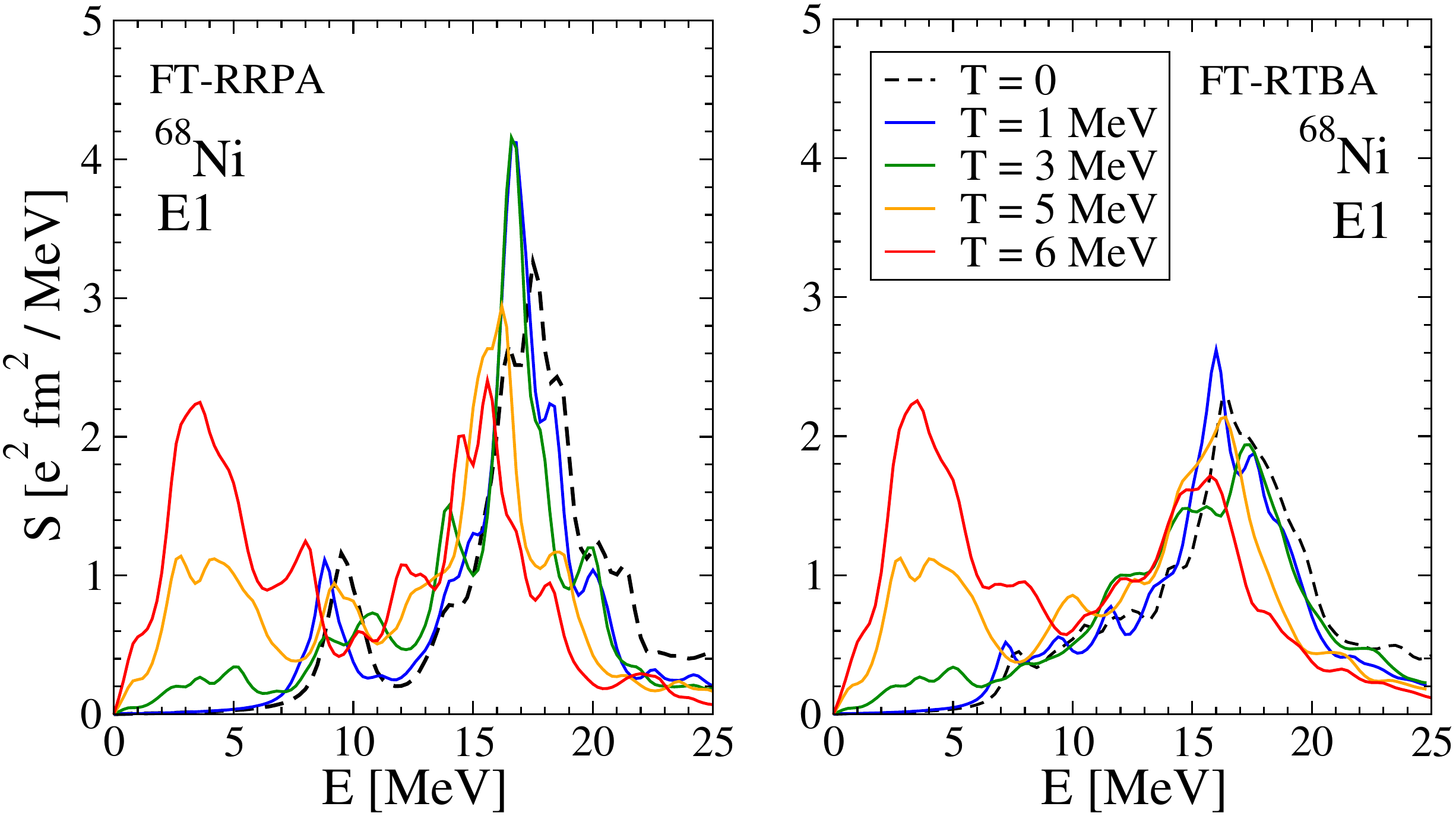}
\caption{Dipole strength distribution in $^{68}$Ni at various temperatures in the finite-temperature relativistic RPA (FT-RRPA) and in the finite-temperature relativistic time blocking approximation (FT-RTBA).}
\vspace{-5mm}
\label{fig-5}       
\end{figure}


Another recent extension of the RNFT was performed for the case of finite temperature \cite{LitvinovaWibowo2018}. For this purpose, the conventional TBA for the time-dependent part of the in-medium nucleon-nucleon interaction amplitude was adopted for the thermal (imaginary-time) Green's function formalism. We found, in particular, that introducing a soft blocking, instead of a sharp blocking at zero temperature, brings the temperature-dependent Bethe-Salpeter equation to a single frequency variable equation similar to Eq. (\ref{Dyson3}). The method was implemented self-consistently in the framework of QHD and investigated the temperature dependence of spectra of even-even medium-heavy nuclei. An illustration of these studies is given in Fig. \ref{fig-5} for the dipole response of $^{68}$Ni. Of particular interest was the spreading of the giant dipole resonance induced by the dynamical kernel of FT-RTBA, which remained strong in the high-temperature regimes, and the appearance of significant low-energy strength caused by the thermal unblocking, see Refs. \cite{LitvinovaWibowo2018,WibowoLitvinova2019} for further details.   The temperature dependence of the spin-isospin response was investigated as well and found to be even more sensitive to low and moderate temperatures \cite{LitvinovaRobinWibowo2018}, that has important implications for astrophysics. 

{\it Summary \textemdash} A novel microscopic approach to the nuclear response developed by synthesis of the equation of motion method and quantum hadrodynamics is presented.
It allows for a consistent description of nuclear excited states with the explicit treatment of beyond-mean-field correlations between up to six fermions, such as $2q\otimes 2phonon$. The first calculations of the dipole response in medium-heavy nuclei demonstrate important improvements of this description in both low- and high-energy sectors.
We also present a finite-temperature version of the theory with explicit $2q\otimes phonon$ correlations. The obtained results are consistent with the existing experimental data and predict the evolution of the nuclear strength functions with temperature taking into account spreading effects. In particular, a noticeable enhancement of the low-lying dipole as well as of the spin-isospin strength distributions has been found at the astrophysical conditions, that may affect the astrophysical modeling of the r-process nucleosynthesis and core-collapse supernovae.
\\
\\
{\it Acknowledgement \textemdash} This work was partly supported by the US-NSF Career Grant PHY-1654379.
%
 \bibliography{Bibliography_Aug2019}
\end{document}